\documentclass[sn-mathphys-num]{sn-jnl}

\usepackage{graphicx}%
\usepackage{multirow}%
\usepackage{amsmath,amssymb,amsfonts}%
\usepackage{amsthm}%
\usepackage{mathrsfs}%
\usepackage[title]{appendix}%
\usepackage{xcolor}%
\usepackage{textcomp}%
\usepackage{manyfoot}%
\usepackage{booktabs}%
\usepackage{algorithm}%
\usepackage{algorithmicx}%
\usepackage{algpseudocode}%
\usepackage{listings}%
\usepackage{bm}

\theoremstyle{thmstyleone}%
\theoremstyle{thmstyletwo}%

\theoremstyle{thmstylethree}%

\begin{document}

\title[Article Title]{Controlled angular momentum injection in a magnetically levitated He II droplet}


\author[1,2]{\fnm{Sosuke} \sur{Inui}}

\author[1,2]{\fnm{Faezeh} \sur{Ahangar}}

\author*[1,2]{\fnm{Wei} \sur{Guo}}
\email{wguo@eng.famu.fsu.edu}

\affil[1]{\orgname{National High Magnetic Field Laboratory}, \orgaddress{\street{1800 East Paul Dirack Drive}, \city{Tallahassee}, \postcode{32310}, \state{Florida}, \country{USA}}}

\affil[2]{\orgdiv{Mechanical Engineering Department}, \orgname{FAMU-FSU College of Engineering, Florida State University}, \orgaddress{ \city{Tallahassee}, \postcode{32310}, \state{Florida}, \country{USA}}}


\abstract{The morphology of rotating viscous classical liquid droplets has been extensively studied and is well understood. However, our understanding of rotating superfluid droplets remains limited. For instance, superfluid $^4$He (He II) can carry angular momentum through two distinct mechanisms: the formation of an array of quantized vortex lines, which induce flows resembling classical solid-body rotation, and surface traveling deformation modes associated with irrotational internal flows. These two mechanisms can result in significantly different droplet morphologies, and it remains unclear how the injected angular momentum is partitioned between them. To investigate this complex problem experimentally, one must first levitate an isolated He II droplet using techniques such as magnetic levitation. However, an outstanding challenge lies in effectively injecting angular momentum into the levitated droplet. In this paper, we describe a magneto-optical cryostat system designed to levitate He II droplets and present the design of a time-dependent, non-axially symmetric electric driving system. Based on our numerical simulations, this system should enable controlled angular momentum injection into the droplet. This study lays the foundation for future investigations into the morphology of rotating He II droplets.}




\maketitle

\section{Introduction}\label{sec: Introduction}
The behavior of rotating classical liquid droplets has been extensively studied, and key insights into the relationship between angular momentum and surface deformation has been established~\cite{Chandrasekhar1965,Brown1980,Lee1998,Hill2008,Baldwin2015}. When a small amount of angular momentum $L$ is imparted, the droplet undergoes solid-body rotation with a constant angular velocity $\Omega$, which increases linearly with $L$ below a critical threshold. Once $L$ exceeds this threshold, the droplet becomes unstable and deforms, typically forming a 2-lobed shape. As $L$ increases further, higher-order deformation modes such as 3-lobed and 4-lobed shapes emerge. Eventually, the solid-body rotation reaches a maximum angular velocity $\Omega_\text{max}$, resulting in a toroidal structure. These higher surface deformation modes are unstable and tend to decay into more stable configurations, such as the 2-lobed shape, or split into smaller droplets. This well-established framework provides a clear connection between angular momentum and surface instabilities in classical fluid dynamics.

However, when it comes to rotating superfluid droplets, such as superfluid helium-4 (He II) droplets, our understanding remains limited. Unlike classical liquid droplets, inviscid superfluid droplets exhibit unique quantum mechanical properties, allowing them to carry angular momentum via two distinct mechanisms. The first involves the formation of an array of quantized vortices, which induce flows inside the droplet resembling classical solid-body rotation. The second mechanism is the excitation of surface traveling deformation modes associated with irrotational flows, giving the appearance of rotation in the droplet’s shape~\cite{Seidel1994}. These two mechanisms are fundamentally different and can result in distinct droplet morphologies and stability. Despite valuable theoretical efforts \cite{Ancilotto2018,Coppens2017,Ancilotto2015,Fetter1974}, a key unanswered question in quantum fluid dynamics is how injected angular momentum is partitioned between these mechanisms and how this partitioning influences the behavior of rotating superfluid droplets.

Several experimental studies have explored the morphology of fast-moving He II nano-droplets ejected into a vacuum pipe through a nozzle~\cite{Gomez2014,OConnell2020,Feinberg2021,Feinberg2022}. By doping these droplets with xenon clusters and using x-ray diffraction imaging, researchers have gained valuable insights into droplet shapes and vortex configurations. However, since each nano-droplet acquires angular momentum in a stochastic manner, these studies lack precise control over the angular momentum injection, and the one-time x-ray snapshots provide only a static view of the droplet profile, not allowing for the observation of droplet dynamics over time. A more promising method for studying He II droplet dynamics is through magnetic levitation. Diamagnetic materials such as water, copper, and even frogs can be levitated in a sufficiently strong, nonuniform magnetic field~\cite{Berry1997}, and this technique has been applied to He II~\cite{Weilert1996,Weilert1997,Vicente2002,Brown2023}. Magnetic levitation offers the advantage of long-term observation of droplet behavior in a stable, unbounded environment. However, despite some early attempts to induce rotation in magnetically levitated He II droplets~\cite{Vicente2002}, the controlled injection of angular momentum into these droplets remains a significant challenge.

In this paper, we describe a magneto-optical cryostat for He II droplet levitation and present our design of a electrical driving system that can generate a time-dependent electric field for controlled angular momentum injection into the levitated droplet. Our numerical simulations demonstrate that this system provides precise control over the droplet’s motion, laying the groundwork for future studies into the morphology of rotating He II droplets. The structure of the paper is as follows: In Sec.~\ref{Setup}, we provide an overview of the cryostat and the magnetic field configuration used to suspend the He II droplets. In Sec.~\ref{sec: Ang Mom Inj}, we describe the system for generating the electric field and how it enables controlled angular momentum injection based on our numerical simulations. Finally, in Sec.~\ref{sec: Disc and Concl}, we summarize our findings and discuss their broader relevance within superfluid helium research.

\section{Magneto-optical cryostat for He II droplet levitation}\label{Setup}
\begin{figure}[t!]
  \centering
  \includegraphics[width=0.8\linewidth]{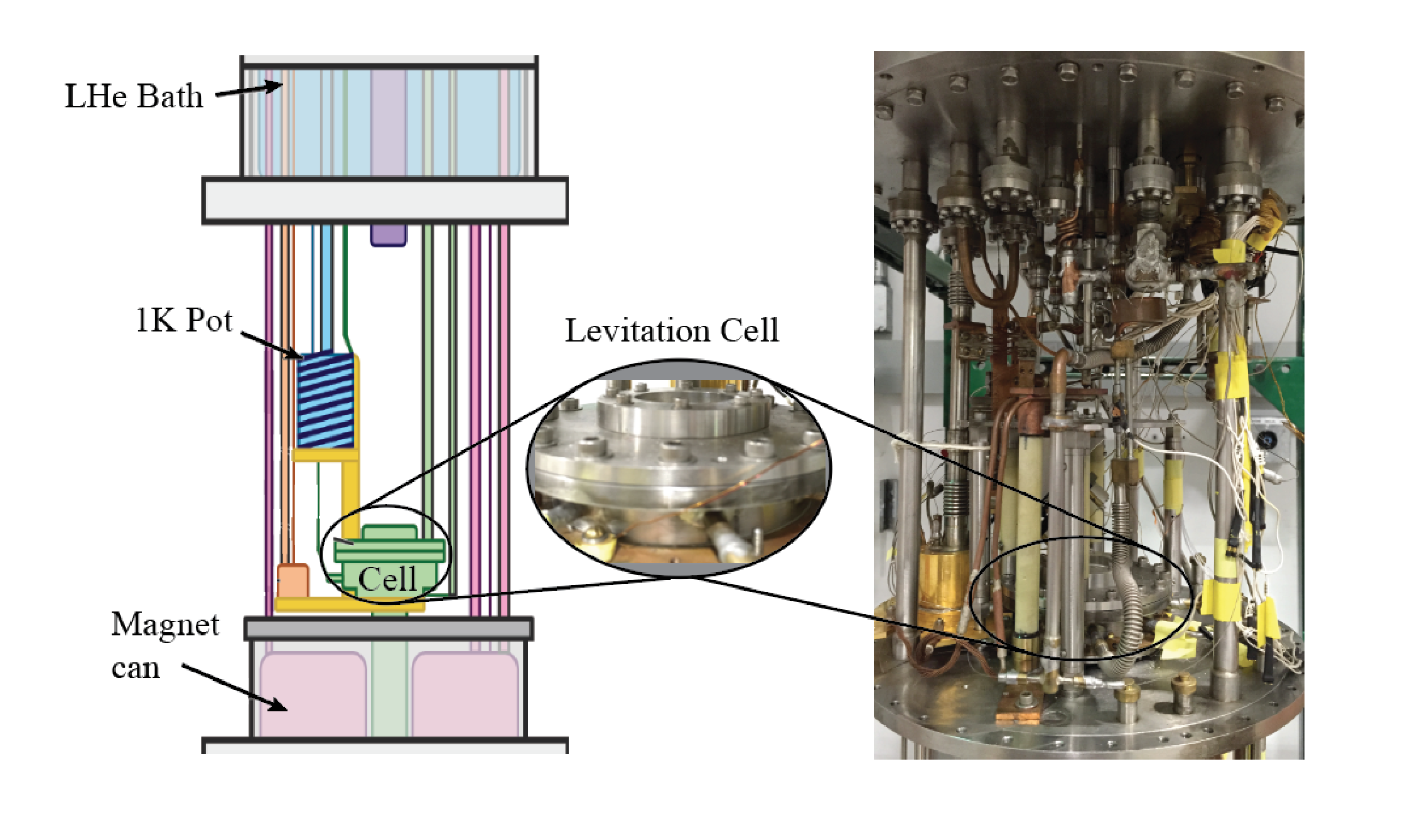}
    \caption{Internal structure of the magneto-optical cryostat for He II droplet levitation. Left: schematic view; Right: photograph of the actual cryostat components. The inset in the center offers a magnified view of the levitation cell. Additional details about the cell are provided in Fig.~\ref{Fig5}.}
    \label{Fig1}
\end{figure}

We use a Janis model 16CNDT cryostat with a custom superconducting magnet to levitate He II droplets. The cryostat, initially designed by Maris' group for their experiments~\cite{Weilert1996,Weilert1997,Vicente2002,Brown2023}, was later transferred to our lab. It comprises four main components: a liquid helium (LHe) bath, a torus-shaped magnet can, a 1-K pot, and an experimental cell. The LHe bath, located in the upper region of the cryostat, is surrounded by a liquid nitrogen shield and a vacuum space to thermally isolate it from the ambient environment. The LHe bath provides cooling power to the 4-K thermal shield, which protects the lower sections of the cryostat from thermal radiation. A continuous flow of LHe from the bath is supplied to the magnet can and the 1-K pot via capillary fill lines. When these components are pumped, they can achieve temperatures as low as 1.3 K. The experimental cell is mounted on the magnet can, with its lower portion extending into the magnet’s central bore, as shown in Fig.~\ref{Fig1}. The cell can be pumped to regulate vapor pressure and, consequently, the He II droplet temperature. Windows installed along the centerline of the cryostat and the cell, aligned with the magnet’s central bore, enable visual observation of the droplet.

\subsection{He II droplet levitation} \label{subsec: Lev mech}
\begin{figure}[t!]
  \centering
  \includegraphics[width=0.6\linewidth]{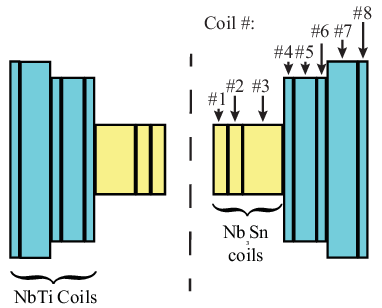}
    \caption{Schematic showing the cross-sectional view of the magnet, consisting of eight concentric superconducting coils made of NbTi and Nb$_3$Sn. The dashed line represents the central axis of all the coils.}
    \label{Fig2}
\end{figure}
The superconducting magnet housed inside the magnet can consists of eight concentric superconducting solenoid coils made of Nb$_3$Sn and NbTi, as depicted in Fig.~\ref{Fig2}. These coils are connected in series, with their dimensions and turn count $N$ provided in Table~\ref{table: Main magnet}.
\begin{table}[h!]
  \centering
  \caption{Configuration of the eight coils that compose the levitation magnet.}\label{table: Main magnet}
  \begin{tabular}{|c|c|c|c|c|c|c|c|c|}
    \hline
    & \textbf{Coil 1} & \textbf{Coil 2} & \textbf{Coil 3} & \textbf{Coil 4} & \textbf{Coil 5} & \textbf{Coil 6} & \textbf{Coil 7} & \textbf{Coil 8} \\
    \hline
    Radius $R$ [mm] & 16 & 22 & 28 & 50 & 56 & 62 & 90 &100  \\
    \hline
    Length $L$ [mm]& 80 & 80 & 80 & 120 & 120 & 120 & 140 & 140 \\
    \hline
    \# of turns $N$ & 960 & 270 & 2150 & 5400 & 5400 & 5400 & 7300 & 7300  \\
    \hline
  \end{tabular}
\end{table}

When a current $I$ is applied, each coil generates an axially symmetric magnetic field. The field $\bm{B}_i(r,z)$ from the $i^\text{th}$ coil is given by:
\begin{equation}
\bm{B}_i(r,z) = \frac{\mu_0  I}{4\pi}  \oint _{\mathcal{L}_i} \frac{\text{d}\bm{l} \times (\bm{r}-\bm{l})} {|\bm{r}-\bm{l}|^3},\label{eq:B}
\end{equation}
where $\bm{l}$ is the position vector along the coil path $\mathcal{L}_i$. If we shift the coordinate to locate the top and bottom faces of the coil at $z = L/2$ and $z = -L/2$, respectively, Eq.~\eqref{eq:B} renders the following expression for the magnetic field components in the $x$-$z$ plane \cite{Derby2010,Caciagli2018}:
\begin{align} \label{eq: B field int}
\begin{split}
B_i^{(x)}(x,z) = &      \frac{-B_0 R}{2} \int_0^{\pi}   \text{d} \phi \cos \phi \\
  &\times  \left[   \frac{1}{ \sqrt{z_{+}^2 +x^2 +R^2 - 2Rx \cos \phi} } -  \frac{1}{ \sqrt{z_{-}^2 +x^2 + R^2 - 2Rx \cos \phi } } \right] \\
B_i^{(y)}(x,z) = &    0  \\
B_i^{(z)}(x,z) = &      \frac{B_0 R}{2} \int_0^{\pi}   \text{d} \phi   \frac {R - x \cos \phi}{x^2 R^2 - 2Rx \cos \phi} \\
  &\times  \left[   \frac{z_{+}}{ \sqrt{z_{+}^2 +x^2 +R^2 - 2Rx \cos \phi} } -  \frac{z_{-}}{ \sqrt{z_{-}^2 +x^2 + R^2 - 2Rx \cos \phi } } \right] \\\end{split}
\end{align}
where $B_0 = \frac{\mu_0 N I}{\pi}$, and $z_{\pm} = z \pm L/2$.
The field $\bm{B}_i(x,z)$ can be calculated by numerically integrating Eq.~\eqref{eq: B field int} using the values of $L$, $R$ and $N$ listed in Table~\ref{table: Main magnet}. As a consequence, the total magnetic field, $\bm{B}(x,z) = \sum_i \bm{B}_i(x,z)$, can be obtained.

\begin{figure}[t!]
  \centering
  \includegraphics[width=1\linewidth]{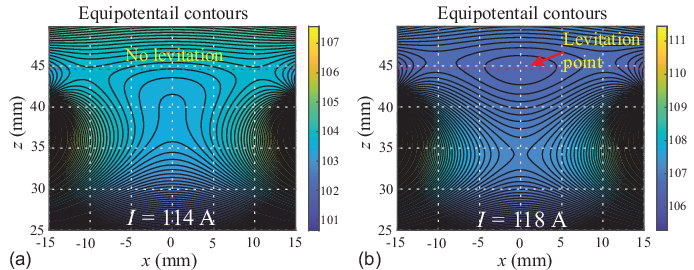}
    \caption{Contour plots of the total potential energy per unit volume $U$ for a LHe droplet placed in the magnetic field produced by the coil system with an applied current of (a) $I = 114$ A and (b) $I = 118$ A, respectively. The location of the levitation point is indicated by a red circle. The quiver plots represent the total force experienced by the droplet.}
    \label{Fig3}
\end{figure}
When a LHe droplet is placed in this nonuniform magnetic field, it experiences a magnetic gradient force, $\bm{F}_m=-\nabla \epsilon$, where $\epsilon = \frac{\chi}{2\mu}\bm{B}^2$ represents the potential energy per unit volume in the magnetic field. Here, $\mu$ is the permeability of LHe, which is close to the vacuum permeability $\mu_0$, and $\chi = -8.6 \times 10^{-7}$ (in SI units) is LHe's magnetic susceptibility~\cite{Wills1924}. To better illustrate the region where the LHe droplet can be stably trapped, we define $U = \epsilon + \rho g z$ as the total potential energy per unit volume of LHe, accounting for the combined effect of the $\bm{B}$ field and gravity, where $\rho$ is the LHe density. Fig.~\ref{Fig3} shows equal-$U$ contour plots for coil currents of 114 A and 118 A. Unlike the case with $I = 114$ A, at 118 A, a trapping region with closed equal-$U$ contours emerges, where $U$ decreases toward the region's center, i.e., the levitation point. At this point, the magnetic force $\bm{F}_m$ balances exactly the gravitational force. The decrease in $U$ toward the levitation point indicates that the total force on the droplet within the trapping region always points toward the levitation point, ensuring stable levitation. The minimum current needed to achieve stable levitation of an LHe droplet is approximately 116 A with this magnet system~\cite{Weilert1996}. When the droplet is slightly displaced from the levitation point in the horizontal plane, the restoring force can be expressed as $F_\text{res} = -kr$, where the spring constant is estimated to be $k \approx 1.8 \times 10^{-5}$ N/m at $I = 118$ A for a droplet with a radius of $a = 1$ mm.  This corresponds to a harmonic oscillation frequency around the levitation point of $f_0 = (2\pi)^{-1}\sqrt{k/M} \approx 0.8 $ Hz, where $M=\rho(4\pi a^3/3)$ is the droplet mass.

\subsection{Droplet Production} \label{subsec: Drop prod}
There are two helium capillary fill lines installed in the cell, both passing through the LHe bath and thermally anchored at the 1-K pot before entering the cell, allowing gaseous helium from an external gas cylinder to condense into the cell at a controlled injection pressure. One fill line exits near the bottom of the cell, while the other exits slightly above the magnetic trapping region, providing two methods for producing a LHe droplet. In the first method, the bottom fill line is used to partially fill the cell, and after the superconducting magnet is activated, helium vapor can spontaneously accumulate and condense in the trapping region, forming a droplet. In the second method, the magnet is turned on first, and helium is condensed from the fill line above the trapping region, with LHe forming at the capillary line's outlet, growing until it detaches and falls into the trapping volume. Once a droplet is successfully levitated, the cell can be pumped to remove vapor, causing the droplet to undergo evaporative cooling, which reduces its size and temperature. The droplet stabilizes at around $T = 0.38$ K, where the vapor pressure becomes negligible (i.e., $P < 10^{-5}$ Torr) and evaporative cooling diminishes. Depending on its initial size, the droplet at this temperature can have a radius between 10$^2$ and 10$^3$ $\mu$m~\cite{Weilert1996,Weilert1997,Vicente2002,Brown2023}.


\section{Angular Momentum Injection}\label{sec: Ang Mom Inj}
\begin{figure}[b!]
  \centering
  \includegraphics[width=1\textwidth]{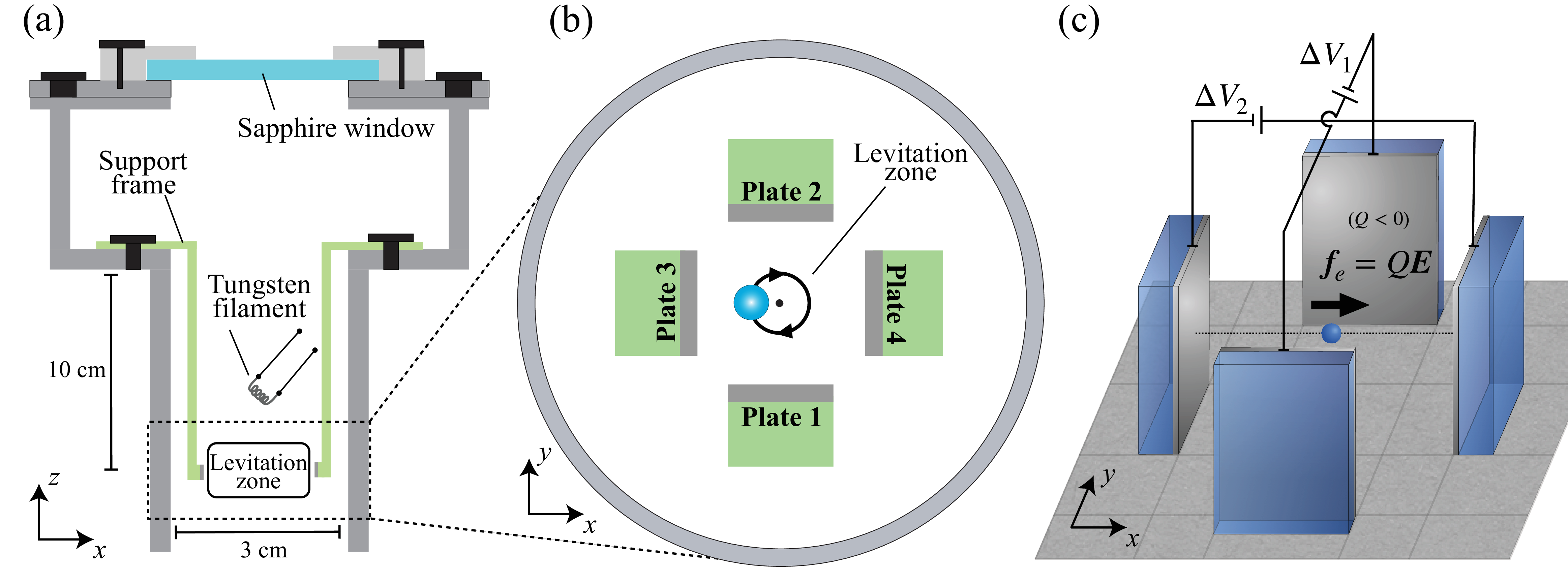}
    \caption{(a) Simplified schematic of the levitation cell (see also Fig. \ref{Fig1}) equipped with a drop driving system, supported by a frame shown in green.  A tungsten filament, depicted above the levitation zone, serves as a thermionic emission source to charge the droplet. (b) Magnified view of the system near the levitation zone in the $xy$-plane.  Each plate has dimensions as specified in Table \ref{table: capacitor design}. (c) Schematic of the conductive plates and voltage supply configuration, where a voltage difference of $\Delta V_1$ ($\Delta V_2$) is applied to Plate 1 and Plate 2 (Plate 3 and Plate 4).}
     \label{Fig4}
\end{figure}

To inject angular momentum into the levitated He II droplet, we propose a procedure comprising four major steps. These steps are outlined below, with details discussed afterward.
\begin{enumerate}
\item \textbf{Charge the droplet:} Charge the droplet surface with electrons generated by thermionic emission from a tungsten filament placed near the droplet.
\item \textbf{Measure the surface charge:} Measure the total charge $Q$ on the droplet surface by balancing the magnetic and electric force.
\item \textbf{Induce orbital motion:} Apply an alternating electric driving force to the levitated droplet using two pairs of plate electrodes oriented in orthogonal directions. Each electrode pair receives suitable periodic voltage pulses to induce an orbital rotation of the droplet around the levitation point, with a radius $R_{orb}$ and angular frequency $f_0$.
\item  \textbf{Convert orbital to spinning motion:} Turn off the driving force once an orbital motion with the desired angular momentum $L$ is achieved. As the orbit radius decreases due to dissipation, the ``orbital'' component of angular momentum is expected to convert into ``spinning'' angular momentum due to conservation of angular momentum.
\end{enumerate}
To inplement these steps, we install an electrical driving system composed of two pairs of parallel conductive plates near the levitation zone and a tungsten filament as an electron source, as depicted in Fig. \ref{Fig4}.  An example set of parameters for the conductive plate design (width $W$, height $H$, and separation $D$) is presented in Table \ref{table: capacitor design}.  Using these parameters,  the electric field generated can be numerically calculated by solving the Poisson equation, $\nabla^2 V = 0$, for the electric potential with appropriate boundary conditions at the plate surfaces and the wall of the computational box.  To estimate $|\bm{E}|$ in Table \ref{table: capacitor design}, we employ the successive over-relaxation (SOR) method \cite{Rapp:2023td}.   The uncertainty in the calculated value represents the standard deviation from the average within the region where the droplet may reside during the angular momentum injection process.
 In the following paragraphs, we discuss each step in detail and describe how the installed driving system acheieves the desired angular momentum injection.

\begin{table}[b!]
\caption{Parameters for the drive system design.  The right-most column shows the electric field near the center when a voltage difference of $\Delta V = 50$ V is applied.}\label{table: capacitor design}
  \centering
  \begin{tabular}{|c|c|c||c|c|}
    \hline
    $W$ [mm] &  $H$ [mm] &  $D$ [mm] &  $\Delta V$ [V]& $|\bm{E}|$ near center [V/cm]\\
    \hline
    $6.0$                      & $20.0$                      & $9.44$                   &        $50.0$             & $49 \pm 2$    \\
    \hline
     \end{tabular}
\end{table}

{\it \underline{Step 1}}:   Charge the droplet surface to drive a levitated droplet efficiently in the horizontal plane.  The surface charge $Q$ must be supplied externally.  A straightforward approach  is via thermionic emission from a tungsten filament placed directly above the droplet, as schematically shown in Fig.~\ref{Fig4} (a).   By applying a voltage difference of a few volts across the filament that is on the order of a few microns in diameter, for a short duration, the filament can thermally emit electrons at a current of several hundreds nA. Due to the strong background magnetic field, the emitted electrons tend to move toward or away from the levitation zone.  If a single pulse of thermionic emission does not achieve the desired charge level, multiple pulses can be applied, ensuring excessive heat is not introduced into the cell.

 The higher the surface charge density $\sigma_e$, the lower the required electric field to drive the droplet.  However, there are limitations to the surface charge density.  One limitation is surface instability when the surface is highly charged, exceeding a critical charge density $\sigma_c$.  Additionally, the surface charge distribution introduces an additional pressure contribution, $P_e = \sigma_e^2 / 2 \epsilon_0$, due to the Coulomb potential.  If $P_e$ exceeds the pressure due to the surface curvature, $P_s = 2 \sigma_s / a$, where $\sigma_s$ is the surface tension, the dynamics of the surface mode may be affected.  Therefore, maintaining the condition $P_s \gg P_e$ is essential to avoid instabilities and ensure stable surface dynamics.

{\it  \underline{Step 2}}:   Evaluate the total amount of charge $Q$ on the droplet.
Since we know the effective spring constant $k$ in the $xy$-plane due to the magnetic trap, $Q$ can be estimated by balancing the magnetic force $f_m$ and electric force $f_e$ generated by a pair of parallel conductive plates (e.g., Plate 1 and Plate 2 in Fig.~\ref{Fig4} (b)).  Experimentally, the following steps should be taken:  Initially, the droplet is at equilibrium.  Gradually increase the voltage difference $\Delta V$ applied between the plates.  The displacement of the droplet, $\Delta x$, will shift proportionally to the force $f_e = QE(\Delta V,W,H,D)$ until equilibrium is restored.  Thus, by measuring  $\Delta x$, the total charge $Q$ can be evaluated as
\begin{equation}
 Q  = \frac{ k \Delta x}{E}.
\end{equation}
If the measured $Q$ is less that the value required in  {\it \underline{Step 3}}, additional electrons may need to be supplied via thermionic emission by repeating the procedure described in {\it \underline{Step 1}}.  Again, care must be taken to avoid introducing excessive heat into the cell during the emission process.

Additionally, the droplet size $a$ should be estimated from captured images at this stage.  Using the obtained values $Q$ and $a$, we can evaluate several key quantities, such as $P_s$ and $P_e$.  If the condition, $P_s \gg P_e$, is not satisfied, the Coulomb interaction may significantly modify the dynamics of the droplet surface modes.   In such cases, an excessively charged droplet should be abandoned , and a new droplet must be prepared.

{\it  \underline{Step 3}}:   Apply voltage signals to two pair of parallel conductive plates, as shown in Fig~\ref{Fig4} (b) and (c), to drive the charged droplet in the levitation zone.  The pairs of orthogonally oriented plates, in principle, allow us to drive the droplet along any orbit within the region enclosed by the plates by modulating the voltages $\Delta V$ across the pairs in a time-dependent manner.
However, since our goal here is to inject angular momentum into the droplet, we focus on a circular orbit with the natural frequency $f_0$ (or $\omega_0 = 2\pi f_0$) of the magnetic trap.

\begin{figure}[b!]
  \centering
  \includegraphics[width=1\linewidth,]{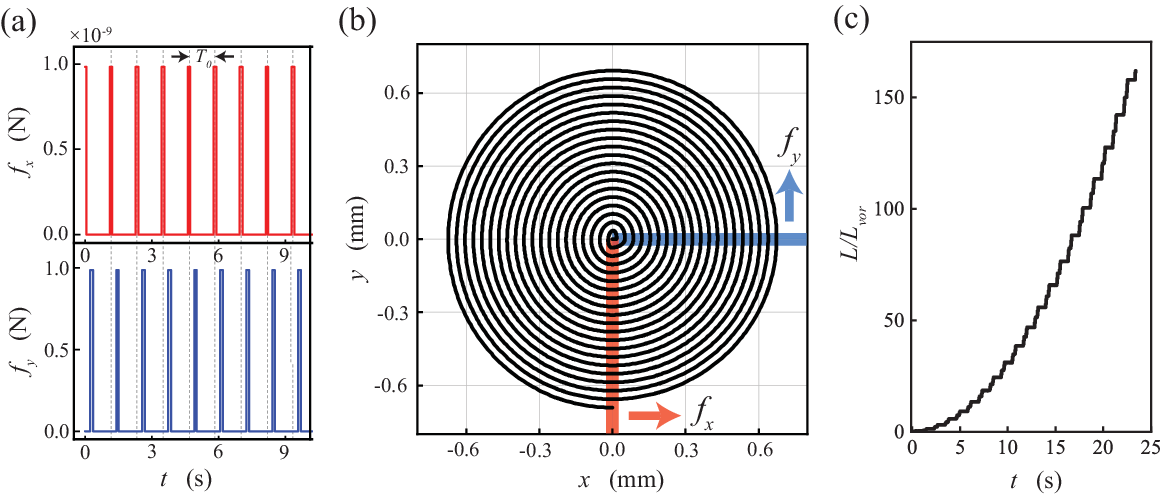}
    \caption{(a) Periodic force on the droplet in the $x$- and $y$-directions due to the voltage pulses applied between the plate pairs.   (b) Simulated orbit of the droplet's center, initially at rest at the origin.  As the pulse signals are applied, the radius of the orbit gradually increases.  The colored regions (red and blue) indicate the locations where the pulse signals ($f_x$ and $f_y$, respectively) are applied.  (c) Dimensionless angular momentum as a function of time.  The normalization factor $L_{vor} = M\kappa/2\pi \approx 9.64\times 10^{-15}$ kg m$^2$/s corresponds to the angular momentum of a droplet with radius $a = 1.0$ mm and a single quantized vortex at its center.}
     \label{Fig5}
\end{figure}

Consider a droplet initially at rest at the potential minimum of the magnetic trap.  We apply voltage pulse signals of constant voltage difference $\Delta V$ and pulse width $\Delta t$ at the natural frequency of the magnetic trap $\omega_0$.  During each cycle, the droplet gains momentum by $\Delta p_x$ and $\Delta p_x$, which can be approximated as $f_e \Delta t$ if $\Delta t \ll T_0$, where $T_0 = 1/f_0$ is the cycle period.   With this momentum, the droplet begins moving along a (nearly) circular orbit at frequency $\omega_0$.  Denoting the instantaneous radius of the orbital motion as $r$, its velocity can be expressed as  $\bm{v} = (-r \omega_0 \sin(\omega_0 t), \omega_0 r \omega_0 \cos(\omega_0 t),0)$.  The efficiency of momentum transfer is maximized when the directions of the electric force $\bm{f}_e$ and $\bm{v}$ are closely aligned.  In the case of optimal momentum injection, the final orbit radius $R_{orb}$ can be estimated in terms of the parameters that characterize the pulse signals: force amplitude $f_e = Q|\bm{E}|$, pulse duration $\Delta t$, and number of pulses $N_\text{pulse}$.  Since $\bm{f}_e \parallel \bm{v}$, the change in linear momentum during each cycle is approximately $f_e \Delta t$.  Furthermore, given that the mass of the droplet $M$ and natural frequency $\omega_0$ remain constant, the relation  $\Delta p \approx  M \omega_0 \Delta R$ holds.  Integrating these expressions over $N_\text{pulse}$ cycles gives:
\begin{equation}   \label{eq: R_orb in terms of others}
 R_{orb} \approx \frac{ f_e N_\text{pulse}\Delta t }{M \omega_0}.
\end{equation}
We observe that $R_{orb}$ is linear to all three parameters ($f_e$, $N_\text{pulse}$, and $\Delta t$).
Consequently, the angular momentum $L = \omega_0 M R_{orb}^2$ scales with the square of these parameters.   Figure~\ref{Fig5} illustrates a numerical simulation result with total surface charge $Q = 2.01 \times 10^{-13}$ C, applied voltage pulses of $\Delta V = 50$ V (corresponding to $f_e = 9.84 \times 10^{-10}$ N) and $\Delta t/T_0 = 0.1$ (see Fig. \ref{Fig5} (a)), for the time duration $0\leq  t \leq 20\times T_0$.     Each $f_x$ and $f_y$ pulse is applied in sync with the natural frequency $\omega_0$, corresponding to the timing when the droplet passes through the region colored in red for $f_x$ and in blue for $f_y$ along the orbit in Fig. \ref{Fig5} (b).  Figure~\ref{Fig5} (c) shows the normalized orbital angular momentum $L/L_{vor}$, where $L_{vor}= M\kappa/2\pi$ represents the angular momentum of a droplet with a single quantized vortex along its center.  This normalization allows the magnitude of $L$ to be interpreted in terms of the number of quantized vortices.  As predicted, $L/L_{vor}$ increases as $N_\text{pulse}^2$.

{\it  \underline{Step 4}}:  Turn off the driving force after $N_\text{pulse}$ cycles, once the desired angular momentum $L$ is achieved.   The droplet will continue rotating about the equilibrium location in a circular orbit for some time.  However, its kinetic energy will gradually dissipate due various processes, such as collisions with background helium gas particles, evaporation of from the droplet surface, and interactions between the surface electrons and the background magnetic field.  Consequently, the orbit radius will gradually diminish. Since the system lacks any obvious mechanism to break axial symmetry, the angular momentum of the droplet is expected to be largely conserved. Consequently, the ``orbital'' angular momentum is anticipated to gradually transform into ``spinning'' angular momentum, carried by surface-traveling deformation modes, flows associated with an array of quantized vortices, or a combination of both.

Although the upper bound of the droplet's final angular momentum is determined by the initial orbital angular momentum, accurately evaluating the final spinning angular momentum, $L_\text{spin}$, is challenging due to uncertainties in the dissipation processes. To precisely determine $L_\text{spin}$, it is essential to measure it independently through other means. Below, we outline two direct measurement approaches (A and B) and one indirect approach (C):

\begin{enumerate}
\item[A.]  Inject fluorescent nano-particles \cite{Meichle2014}.  By tracking of the motion of such particles, the velocity field within the droplet can be reconstructed.  The fluorescent particle may be introduced during the droplet formation process.  Liquid helium is sent to the levitation cell through a capillary.  Thus,  by coating the capillary wall with the nano particles in prior to the droplet formation, we could effectively introduce them in the droplet.\\
\item[B.] Inject nano- to micron-sized metallic particles produced by laser ablation of a metallic target place near the droplet \cite{Moroshkin2010}.  Those particles can act as tracers to visualize the flow field or quantized vortices. \\
\item[C.] Measure the surface distortion.  If $L/L_{vor} \gg 1$, the behavior of the droplet may mimic that of a classical droplet.  Then, the angular momentum associated with the spinning motion may be evaluated from the deformation of the droplet.   \\
\end{enumerate}
Approach C could be a powerful tool to evaluate $L_\text{spin}$. However, since quantized vortices and irrotational flow may coexit within the He II droplet, it is not guaranteed that the classical observations apply directly to this system.   Once direct measurements (A or B) are conducted, the performance of the approach C can be assessed systematically by investigating the relationship between deformation and angular momentum.


\section{Summary}\label{sec: Disc and Concl}
In this paper, we have presented a detailed analysis of the magnetic levitation of liquid helium droplets using a superconducting coil magnet and explored the associated phenomena. We described the structure of our magneto-optical cryostat and explained the mechanism that enables stable levitation of helium droplets, providing a unique platform for investigating the dynamics of quantum fluids. A key focus of our work is the surface deformation of a helium droplet when it rotates with a given angular momentum $L$. While surface deformation and angular momentum transfer in classical fluids are well understood, their counterparts in superfluid helium (He II) remain largely unexplored. Unlike a classical droplet, where angular momentum is associated with solid body rotation, in He II, as highlighted in Ref.~\cite{Seidel1994}, part of $L$ may be carried by the irrotational component of the bulk flow. To address these questions, we designed an experimental setup capable of injecting angular momentum into a levitated helium droplet. By inducing steady orbiting motion and subsequently turning off the driving force, the orbital angular momentum is expected to largely convert into spinning angular momentum. This setup allows for precise control of $L$ and the observation of surface deformations caused by the bulk flow. By comparing the results with classical fluid dynamics, we aim to disentangle the contributions of quantized vortices ($L_\text{vor}$) and irrotational flow ($L_\text{irr}$) to the total angular momentum. These studies could have broader implications for understanding quantum fluid dynamics and may provide insights into astrophysical phenomena, such as glitches in neutron stars \cite{Ginzburg1964,Sauls1989}.

\backmatter



\section*{Acknowledgments}
The authors acknowledge the support by the National Science Foundation under Award No. OSI-2426768 and the Gordon and Betty Moore Foundation through Grant No. GBMF11567. The work was conducted at the National High Magnetic Field Laboratory at Florida State University, which is supported by the National Science Foundation Cooperative Agreement No. DMR-2128556 and the state of Florida.









\bibliography{DLC}

\end{document}